# Emerging Security Challenges of Cloud Virtual Infrastructure

Amani S. Ibrahim, James Hamlyn-Harris and John Grundy
*Computer Science & Software Engineering, Faculty of Information & Communication Technologies*
*Swinburne University of Technology, Hawthorn, Victoria, Australia*
[aibrahim, JHamlynharris, jgrundy]@ swin.edu.au

*Abstract* — **The cloud computing model is rapidly transforming the IT landscape. Cloud computing is a new computing paradigm that delivers computing resources as a set of reliable and scalable internet-based services allowing customers to remotely run and manage these services. Infrastructure-as-a-service (IaaS) is one of the popular cloud computing services. IaaS allows customers to increase their computing resources on the fly without investing in new hardware. IaaS adapts virtualization to enable on-demand access to a pool of virtual computing resources. Although there are great benefits to be gained from cloud computing, cloud computing also enables new categories of threats to be introduced. These threats are a result of the cloud virtual infrastructure complexity created by the adoption of the virtualization technology.**

**Breaching the security of any component in the cloud virtual infrastructure significantly impacts on the security of other components and consequently affects the overall system security. This paper explores the security problem of the cloud platform virtual infrastructure identifying the existing security threats and the complexities of this virtual infrastructure. The paper also discusses the existing security approaches to secure the cloud virtual infrastructure and their drawbacks. Finally, we propose and explore some key research challenges of implementing new virtualization-aware security solutions that can provide the pre-emptive protection for complex and ever-dynamic cloud virtual infrastructure.**

I. INTRODUCTION

Cloud Computing [1, 2] is a new computing paradigm in which the Internet is used to deliver reliable IT services to customers. The amount of that service can be scaled up and down based on customer needs. This flexibility, combined with the potential of a "pay-per-use" model makes the cloud attractive solution to enterprises, where the capital expenses are heavily reduced. Cloud Computing is a combination of existing technologies that make a paradigm shift in building and maintaining distributed computing systems. The large improvements in processors, virtualization technology, data storage and networking have combined to make the cloud computing a more compelling paradigm. The cloud computing service model is "X-as-a-Service", where X includes IT functions (e.g. infrastructure, storage, platform, database, software, security).

Infrastructure-as-a-Service (IaaS) [3] is one of most popular and important services delivered by the cloud computing model. IaaS allows customers to increase their available computational and storage resources *on the fly* without investing in their own hardware.

IaaS is characterized by the concept of resource virtualization, which is a key enabler of enterprise cloud computing. Virtualization technology enables the execution of multiple operating system instances - called virtual machines (VMs) - on the same physical server. Each VM functions as if it is the only owner of the physical server with a dedicated operating system and hosted applications. Virtualization technology provides the capability to achieve higher hardware utilization rates and cut costs by aggregating a collection of physical servers into one server. The virtual infrastructure inside a cloud physical server is composed from three core components, as shown in Figure 1.

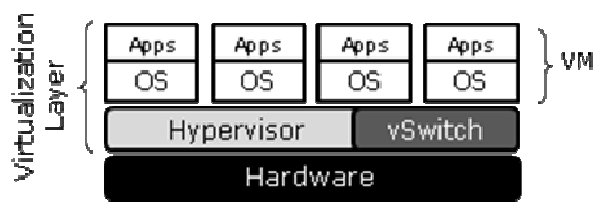

**Figure 1. Server Virtualization in Cloud Computing**

1) **Hypervisor** - The hypervisor acts as the abstraction layer that provides the necessary resource management functions that enable sharing of hardware resources between the VMs. Hypervisors have two main models: hosted (para-virtualization), such as Xen and Hyper-V, and non-hosted (full-virtualization), such as VMware. These two models trade-off some level of isolation to increase sharing of resources among VMs. Typically, isolation comes at the cost of performance.

2) **Virtual Network** - the virtual network contains the virtual switch (vSwitch) software that controls multiplexing traffic between the Virtual NICs (VNICs) of the installed VMs and Physical NICs (PNICs) of the physical host. The vSwitch also controls the inter-VM traffic on a single host that doesn't touch the PNICs of the host, and manage the customers trust zones. The vSwitch acts like a physical switch in a non-virtualized environments, and nearly do the same tasks, such as the core layer 2 forwarding functions, VLAN tagging, layer 2 checksum and segmentation. However, some functions like Spanning Tree Protocol are not needed in the vSwitch because there is no way to make redundant switch connections.

3) **Virtual Machines (VMs)** - VMs are the software entities that emulate a real physical machine. VMs run under the control of the hypervisor that virtualize and multiplexes the hardware resources.

The reset of this paper is organized as follows. Section 2 explores the cloud virtual infrastructure security problems and the different threats that can affect the virtual infrastructure components. Section 3 reviews the previous work in the area of securing the virtual cloud infrastructure and virtualized servers. In section 4, we explore the key research challenges of implementing security solutions to protect the cloud virtual infrastructure. Finally, section 6 concludes the paper with a summary of its research contribution.

## II. CLOUD VIRTUAL INFRASTRUCTURE SECURITY

Cloud computing model provides organizations with a more efficient, flexible and cost effective alternative to own their computing resources. However, hackers and security researchers have shown that these capabilities of virtualization can be exploited to create new and more robust forms of malware that are hard to detect and can evade current security technologies [4].

### A. Threat Model

Security responsibility in the cloud is not a single-side responsibility. Security is shared between the cloud provider and the cloud user. Customers are not aware of how their VMs are being protected. On the other hand, the cloud providers running VMs are not aware of the VM contents. Thus, there is no complete trust relationship between cloud customers and providers. From a cloud provider perspective, customers' VMs cannot be trusted and this will be our research focus. In our threat model, a hacker can be cloud user that hosts a service or non-cloud user, and in both models the victim is the cloud provider that runs the service or the other hosted VMs. In the former threat model, hackers have more chances of success, because they have access to the Cloud Virtual Infrastructure (VCI), and can run different malware to gain more access privileges.

### B. Security Threats

Breaching the security of any component in the VCI impacts significantly on the security of the other components and consequently affects the overall system security. In these papers [5-7], the authors investigated different vulnerabilities and security threats in cloud computing focusing on the VCI security threats. Security threats for the cloud virtual infrastructure can be divided into three categories:

1) **Hypervisor Attacks** - Hackers consider the hypervisor a potential target because of the greater control afforded by lower layers in the system. Compromising the hypervisor enables gaining control over the installed VMs, the physical system and hosted applications. HyperJacking [8, 9], BLUEPILL [10], Vitriol [11], SubVir [12] and DKSM [13] are well-known attacks that target the virtual layer at run-time. These VM-Based Rootkits (VMBRs) are capable of inserting a malicious hypervisor *on the fly* or modifying the installed hypervisor to gain control over the host workload. In some hypervisors like Xen [14], the hypervisor is not alone in administering the VMs. A special privileged VM serves as an administrative interface to Xen, and control the other VMs. This VM is also a potential target for hackers target to exploit vulnerabilities inside that VM to gain access to the hypervisor or the other installed VMs.

1) **vSwitch Attacks** - The vSwitch is vulnerable to a wide range of layer-2 attacks like a physical switch. These attacks include vSwitch configurations, VLANs and trust zones, and ARP tables [15].

2) **Virtual Machine Attacks** - Cloud servers contain tens of VMs, these VMs may be active or offline, and in both states they are vulnerable to various attacks. Active VMs are vulnerable to all traditional attacks that can affect physical servers. Once a VM is compromised, this gives the VMs on the same physical server a possibility of being able to attack each other, because the VMs share the same hardware and software resources e.g. memory, device drivers, storage, hypervisor software. Co-location of multiple VMs in a single server and sharing the same resources, increases the attack surface and the risk of VM-to-VM or VM-to-hypervisor compromise [16]. On the other hand, when a physical server is off, it is safe from attacks. However, with VMs when a VM becomes offline, it is still available as VM image files that are susceptible to malware infections and patching. Additionally, provisioning tools and VM templates are exposed to different attacks that target to create new unauthorized VMs, or patch the VM templates to infect the other VMs that will be cloned from this template.

These new categories of security threats are a result of the new, complex and dynamic nature of the cloud virtual infrastructure, as follows:

- **Multi-Tenancy -** Different users within a cloud share the same applications and the physical hardware to run their VMs. This sharing can enable information leakage exploitation and increases the attack surface and the risk of VM-to-VM or VM-to-hypervisor compromise.
- **Workload Complexity -** Server aggregation duplicate the amount of workload and network traffic that runs inside the cloud physical servers, which increase the complexity of managing the cloud workload.
- **Loss of Control -** users are not aware of the location of their data and services and the cloud providers run VMs they are not aware of their contents.
- **Network Topology -** The cloud architecture is very dynamic and the existing workload change over time, because of creating and removing VMs. In addition, the mobile nature of the VMs that allows VMs to migrate from one server to another leads to non-predefined network topology.
- **No Physical Endpoints -** Due to server and network virtualization, the number of physical endpoints (e.g. switches, servers, NICs) is reduced. These physical endpoints are traditionally used in defining, managing and protecting IT assets.
- **Single Point of Access -** virtualized servers have a limited number of access points (NICs) available to all VMs. This represents a critical security vulnerability where compromising these access points opens the door to compromise the VCI including VMs, hypervisor or the vSwitch.

## III. RELATED WORK

The virtualization security research area was a concern even before the cloud computing era. Research in this area can be categorized into: Traditional Security Solutions in the Cloud, Virtualization-Aware Security Solutions, Micro Hypervisors; and Hypervisor-Level Protection. These different problem areas are explored in the following sub-sections.

### A. Cloud Computing and Traditional Security Solutions

This category of research focuses on how to use current security technologies, including firewalls, IDSs and IPSs, to secure the cloud virtual infrastructure. Fagui et al [17] use a firewall to protect the Xen hypervisor virtual network. This framework is based on a firewall hook framework called "Netfilter/Iptables" [18]. Sebastian et al [19] introduced a conceptual cloud-based IDS deployment model. This model is based on deploying IDS sensors with each layer with a centralized IDS management module. Kleber et al [20] explored IDS as security software for the cloud by applying behavioral and knowledge-based analysis techniques to detect known and unknown attacks. Amir et al [21] applied agent-based IDS as a security solution for the cloud. Jia et al [22] introduced a framework that is based on the network-based IPS to install network filters in the cloud. Security approaches that rely on deploying traditional security solutions in the VMs to secure the CVI cloud virtual infrastructure have significant limitations. These approaches have a significant performance impact on the system as they generally need to trap every system call, I/O request and memory access before forwarding it to the hypervisor. They also cannot prevent attacks between VMs and the vSwitch because this approach does not leverage hypervisor-aware security capabilities. Moreover, these approaches are used regardless of the cloud complexity that results from the unlimited number of changes in the cloud topology, VM mobility and dynamic states, the huge number of the monitored objects and network traffic, and the inter-VM communications. On the other hand, as host-based security solutions install security agents or drivers on each VM to perform monitoring, these agents can be detected by the new-generation rootkits that have the ability to detect the installed security software and tamper with its behavior. Another important missing area in current research is the security of the vSwitch software and VLAN configurations.

### B. Virtualization-Aware Security Solutions

This security approach deploys the security software in a dedicated and privileged VM (SecVM) with privileged access to the hypervisor to secure the other VMs (untrusted VMs) installed in the same physical server, as shown in Figure 2.

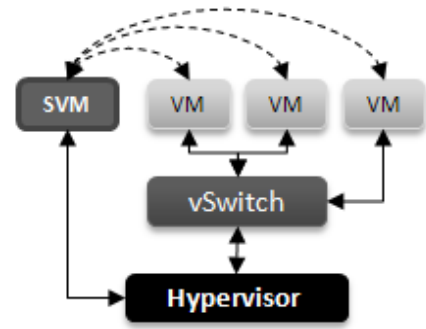

**Figure 2. SecVM and VMI Security Approach.**

The SecVM utilizes Virtual Machine Introspection (VMI) techniques, to enable monitoring and observing VMs from outside a VM, and get a view of the VM at the hypervisor level. This approach makes use of the isolation feature of virtualization to ensure that the security solution is isolated from the other server workload, and is also installed in a layer lower than the one being protected. VMI is used to monitor untrusted VMs from outside the VM itself without installing any hooks or drivers inside the untrusted VM. This approach makes it harder for hackers to detect the installed security software and the SecVM becomes protected from any attack target tampering with the security software.

VMI and SecVM model research has gained noticeable attention. VMwatcher [23], VMwall [24], and others were developed to monitor the VM from a hypervisor perspective. A major drawback of the previous researches is the loss of semantic information. From outside the VM we get a VM view from a hypervisor perspective which includes memory-pages, disk-blocks and low-level instructions. In contrast when monitoring from inside the VM, we can view high-level entities like processes, registry keys, files, events, traffic and system calls. The difference between outside and inside views is called the semantic gap. X-Spy [25] is an IDS, which makes use of the VMI and SecVM approach to build a security solution that overcomes the semantic gap problem. Lares [26], and VMSec [27] overcome the semantic gap problem by taking into consideration the kernel version structure.

By leveraging virtualization-aware security software, enterprises can enable different security technologies across all VMs on a protected physical server. However some security functions, such as handling encrypted traffic, accessing certain real-time information or the process of cleaning and removing malware from infected VMs will continue to require VM-based agents. While introspection has many applications, it is fundamentally limited because it only can perform passive monitoring [26]. Thus introspection is not sufficient for security applications that rely on active monitoring. Also, these approaches don't consider the dynamic and mobile nature of VMs and cloud components, and only provide security for a limited number of VMs inside a physical host.

### C. Micro Hypervisors

This category of research aims to develop new secure hypervisors with a specialized micro-kernel instead of the current large foot-print hypervisors. Micro hypervisors include the necessary abstractions and management functions in the kernel mode (Ring 0), and their other functions and the device drivers are developed in an upper layer (Ring 1 or user-mode). This approach helps to shrink the most critical attack surface in the hypervisor according to the Trusted Computing Base (TCB) rule [28]. Takahiro et al [29] and Udo et al [30] have developed examples of micro-hypervisor implementations. This category is out of our current research scope as our research is focusing on solutions to secure current heavy-weight hypervisors.

### D. Hypervisor-Level Protection

This category of security research aims to secure the hypervisor itself against hypervisor-based rootkits and page-level memory attacks that arise from shared memory pages and software-level memory page translations. Ryan [31] provide memory management techniques and secure software-based page-level protection to secure the hypervisor. However, new generations of the AMD Opteron and Intel Xeon series processors have provided hardware support for memory virtualization by using two layers of page tables in the hardware-level instead on software-level translations. This increases the trust of the hardware and the hypervisor and restricts the boundaries of access that a VM can have. Intel and AMD virtualization technology hardware also provides powerful features to enhance trusting and protecting the operating platform, these features include multi-queue network, I/O memory management and isolation enforcements, directed I/O. These features help the hypervisor to be more robust and restrict the range of physical memory locations that I/O devices are able to access [32].

## IV. KEY RESEARCH CHALLENGES

The cloud virtual infrastructure is very complex and dynamic. In addition, the huge amount of traffic and workload flowing inside each physical server increases the complexity of the protected environment. The virtual architecture of the cloud erases many of the physical boundaries that are traditionally used in defining, managing and protecting organizations' IT assets, leading to a very complex virtual architecture. Adapting security solutions in the cloud environment to protect cloud virtual infrastructure is a real challenge and requires key characteristics to be addressed in order to deliver the accurate and pre-emptive protection. These key characteristics include:

*1)* **Performance** – Running security software impacts performance where the security software typically needs to trap e.g. system calls, I/O request and memory access before forwarding it to the hypervisor. Trapping every system activity within the huge amount of cloud platform activities is a major challenge. Moreover, providing real-time monitoring for the CVI including VM workload, vSwitch, and hypervisor is a challenging task in such complex and dynamic environment. On the other hand, more than just real-time monitoring is needed. A cloud security system should ideally have the intelligence to be self-defending and be able to prevent threats, not just detect. To achieve that, active monitoring should be implemented; not passive monitoring. Active monitoring means installing monitors that can suspend system activities and events from execution until the event is being inspected by the security software, but in passive monitoring, monitors can see just the events but cannot interrupt these events. In passive monitoring; the security system cannot stop threats even if the monitoring is occurring in real-time.

**2) Zero-day Threats Detection** – The ability to detect zero-day (unknown) threats is a goal for security experts. Security threats increase every day in number, complexity and creativity, so the hacker behaviour can't be easily predicted. Focusing on threat behaviour may cause a huge number of false positives affecting the protected system with a range of zero-day threats. Instead, behavioural analysis techniques that focus on the protected system behaviour rather than the threat behaviour are a potentially effective approaches that enable detecting the zero-day threats with a low rate of false positives and negatives [33]. A good picture of cloud behaviour can be developed by monitoring different components and activates inside the cloud. However, maintaining real-time monitoring for the large number of cloud activities and components makes it more difficult to build a model for the ideal normal behaviour file of the cloud. To build an accurate protection system on the basis of monitoring cloud behaviour we must carefully select the critical objects in the cloud that are targets for hackers and can lead to useful monitoring results.

**3) Control** – A major goal of hackers is control, by which hackers will have the ability to monitor, intercept, and modify system events and activities. Control of a system is determined by which side occupies the lower layers in the software stack, where lower layers control upper layers because lower layers implement the abstractions upon which upper layers depend. Controlling the system allows malware to remain invisible by obviating or disabling the security software. Virtualization -aware security software should be installed in the lowest layer of the software stack of the whole cloud platform, not only the software stack of the VM (VM's OS is not the lowest layer in the virtualized environments). The security software should be implemented starting from the hypervisor.

**4) Defence-in-Depth** – having a defence-in-depth approach is fundamental for providing a trustworthy cloud infrastructure [33, 34]. Defence-in-depth means defending the cloud virtual infrastructure at different layers with different protection mechanisms, according to the layer characteristics. Applying such a defence strategy ensures that threats should bypass by one or more of the defence layers. This strategy enables identifcation and blocking of threats at early stages before they propagate into the cloud workload.

**5) Virtual Appliance and JeOS** –With the advent of virtualization, the industry is in need of a new software delivery system that leverages all the benefits of virtual infrastructures. The current approach to software delivery is costly and complex, especially when it comes to enterprise as hardware-based appliances. Virtual appliances offer a new paradigm for software delivery by packaging pre-configured, virtualization-ready solutions in a single software package that is secure, easy to distribute, and easy to manage [35]. A virtual appliance is a preconfigured software solution running on a pre-configure virtual machine with just enough operating system (JeOS) - purpose-built operating systems - that supports only the functions of the application. JeOS solutions occupy a much smaller footprint than general-purpose operating systems and are more stable and secure because they contain fewer lines of code, reducing the number of vulnerability exploits or configuration conflicts which can occur.

**6) Monitored VM** – Providing an isolated, weakened or unprotected VM hosting mock services, which is carefully monitored and constrained, may enable detection of emerging threats by monitoring this VM behaviour. Hackers will detect and attempt to compromise this VM, revealing attack strategies that can be counteracted for other VMs hosting real application services. Such VMs should be carefully installed and managed, as they may become a foothold for attacks from within the cloud infrastructure.

## V. CONCLUSION

There are security challenges in the cloud, and a secure cloud is impossible unless the virtual environment is secure. Traditional security solutions do not map well to the virtualized environments, because of the complex and ever-dynamic nature of the cloud computing. New virtualization-aware security solutions should be provided to ensure the preemptive security to the overall system. These security solutions should have the intelligence to be self-defending and have the ability to provide real-time detection and prevention of known and unknown threats.

Our research is focusing on developing a new virtualization-aware security solution that can meet our research challenges and have the ability to defend the cloud virtual infrastructure different layers (including VMs, vSwitch and Hypervisor) against zero-day threats.

## REFERENCES
[1] Alexander Lenk, Markus Klems, Jens Nimis, Stefan Tai, and Thomas Sandholm, "What's inside the Cloud? An architectural map of the Cloud landscape," in *Proceedings of the 2009 ICSE Workshop on Software Engineering Challenges of Cloud Computing*, 2009, pp. 23-31.

[2] Luis Vaquero, Luis Rodero-Merino, Juan Caceres, et al, "A break in the clouds: towards a cloud definition," *ACM SIGCOMM Computer Communication Review,* vol. 39, pp. 50-55, 2009.

[3] Wesam Dawoud, Ibrahim Takouna and Christoph Meinel, "Infrastructure as a service security: Challenges and solutions," in *2010 The 7th International Conference on Informatics and Systems*, 2010, pp. 1-8.

[4] Kevin Skapinetz, "Virtualisation as a Blackhat Tool," in *Network Security, Elsevier.*, 2007, pp. 4-7.

[5] W. Dawoud, , Takouna, I., Meinel, C., "Infrastructure as a service security: Challenges and solutions," in *he 7th International Conference on Informatics and Systems*, Cairo, May 2010.

[6] Kai Hwang, Sameer Kulkareni, Yue Hu, "Cloud Security with Virtualized Defense and Reputation-Based Trust Mangement," Eighth IEEE International Conference on Dependable, Autonomic and Secure Computing, 2009, pp.717-722.

[7] Bernd Grobauer, Tobias Walloschek and Elmar Stöcker, "Understanding Cloud-Computing Vulnerabilities," IEEE Security and Privacy, 10 Jun. 2010. IEEE computer Society Digital Library. IEEE Computer Society, pp.1-8.

[8] Martim Carbone, Diego Zamboni, Wenke Lee, "Taming Virtualization," IEEE Security and Privacy, 2008, vol. 6, pp. 65-67.

[9] Edward Ray, and Eugene Schultz, "Virtualization security," in *Proceedings of the 5th Annual Workshop on Cyber Security and Information Intelligence Research: Cyber Security and Information Intelligence Challenges and Strategies*, Oak Ridge, Tennessee, 2009, pp. 1-5.

[10] Joanna Rutkowska, "Subverting VistaTM Kernel for Fun and Profit, ," *Black Hat Conference,* 2006.

[11] Dino Dai Zovi, "Hardware Virtualization Rootkits," in *BlackHat Conference*, USA, 2009.

[12] Samuel King, Peter Chen, Yi-Min Wang, et al, "SubVirt: Implementing malware with virtual machines," 2006 IEEE Symposium on Security and Privacy, 2006, pp.314-327.

[13] Sina Bahram, Xuxian Jiang, Zhi Wang, Mike Grace, et al, "DKSM: Subverting Virtual Machine Introspection for Fun and Profit," Proceedings ofthe 29th IEEE International Symposium on Reliable Distributed Systems, New Delhi, India, October 2010.

[14] Xen Open Source, "How Does Xen Work?," in *http://www.xen.org/files/Marketing/HowDoesXenWork.pdf*, Access on July 2010, 2009.

[15] Serdar Cabuk, Chris Dalton, Aled Edwards, et al, "A Comparative Study on Secure Network Virtualization," in *Technical Report No. HPL-2008-57, HP Labs*, 2008, http://www.hpl.hp.com/techreports/2008/HPL-2008-57.pdf, Accessed on June 2010.

[16] Tavis Ormandy, Google, "An Empirical Study into the Security Exposure to Host of Hostile Virtualized Environments. ," in *Applied Security Conference*, Vancouver, British Columbia, 2007.

[17] Fagui Liu, Xiang Su, Wenqian Liu, et al, "The Design and Application of Xen-based Host System Firewall and its Extension," in *The 2009 International Conference on Electronic Computer Technology*, 2009, pp. 392-395.

[18] "Netfilter project," *2008,* http://www.netfilter.org/, Accessed July 2010.

[19] Sebastian Roschke, Feng Cheng and Christoph Meinel, "Intrusion Detection in the Cloud,"Eighth IEEE International Conference on Dependable, Autonomic and Secure Computing, 2009, pp.729-734.

[20] Kleber Vieira, Alexandre Schulter, Carlos Westphall, et al "Intrusion Detection for Grid and Cloud Computing," IT Professional, July/Aug. 2010, vol. 12, pp. 38-43.

[21] Amir Dastjerdi, Kamalrulnizam Abu Bakar, and Sayed Tabatabaei, "Distributed Intrusion Detection in Clouds Using Mobile Agents," in *Proceedings of the 2009 Third International Conference on Advanced Engineering Computing and Applications in Sciences*, 2009, pp. 175-180.

[22] Jia Tiejun, and Wang Xiaogang, "The Construction and Realization of the Intelligent NIPS Based on the Cloud Security," in *1st International Conference on Information Science and Engineering (ICISE)*, Nanjing 2009, pp. 1885 - 1888.

[23] Xuxian Jiang, Xinyuan Wang, Dongyan Xu, "Stealthy malware detection through vmm-based "out-of-the-box" semantic view reconstruction," in *Proceedings of the 14th ACM conference on Computer and communications security*, Alexandria, Virginia, USA, 2007, pp. 128-138.

[24] Abhinav Srivastava, and Jonathon Giffin, "Tamper-Resistant, Application-Aware Blocking of Malicious Network Connections," in *Proceedings of the 11th international symposium on Recent Advances in Intrusion Detection*, Cambridge, MA, USA, 2008, pp. 39-58.

[25] Bernhard Jansen, Harigovind Ramasamy, Matthias Schunter, et al, "Architecting Dependable and Secure Systems Using Virtualization " in *Architecting Dependable Systems V*, 2008, pp. 124-149.

[26] Bryan D. Payne, Martim Carbone, Monirul Sharif, Wenke Lee, "Lares: An Architecture for Secure Active Monitoring Using Virtualization," 2008 IEEE Symposium on Security and Privacy (sp 2008), 2008, pp.233-247.

[27] Flavio Lombardi, and Roberto Di Pietro, "KvmSec: a security extension for Linux kernel virtual machines," in *Proceedings of the 2009 ACM symposium on Applied Computing*, Honolulu, Hawaii, 2009, pp. 2029-2034.

[28] Lenin Singaravelu, Calton Pu, Hermann Hrtig, Christian Helmuth, "Reducing TCB complexity for security-sensitive applications: three case studies," *SIGOPS Oper. Syst. Rev.,* vol. 40, pp. 161-174, 2006.

[29] Takahiro Shinagawa, Hideki Eiraku, Kouichi Tanimoto, et al, "BitVisor: a thin hypervisor for enforcing i/o device security," presented at the Proceedings of the 2009 ACM SIGPLAN/SIGOPS international conference on Virtual execution environments, Washington, DC, USA, 2009.

[30] Udo Steinberg, Bernhard Kauer, "NOVA: a microhypervisor-based secure virtualization architecture," in *Proceedings of the 5th European conference on Computer systems*, Paris, France, 2010, pp. 209-222.

[31] Ryan Riley, Xuxian and Dongyan Xu, "Guest-Transparent Prevention of Kernel Rootkits with VMM-Based Memory Shadowing," in the Proceedings of the 11th international symposium on Recent Advances in Intrusion Detection, Cambridge, MA, USA, 2008.

[32] Rich Uhlig, Alberto Munoz, "Resource Protection in Virtualized Infrastructures," *Intel® Virtualization Technology,* August 2009.

[33] Amani Salah Ibrahim, Mohamed Shouman, Hossam Faheem, "Surviving cyber warfare with a hybrid multiagent-base intrusion prevention system," *Potentials, IEEE,* vol. 29, pp. 32-40, 2010.

[34] Microsoft Research, "Securing Microsoft's Cloud Infrastructure," in *White Paper*, 2009, http://www.globalfoundationservices.com/security/documents/SecuringtheMSCloudMay09.pdf, Accessed on August 2010.

[35] VMware, "Virtual Appliances: A New Paradigm for Software Delivery, http://www.vmware.com/files/pdf/vam/VMware_Virtual_Appliance_Solutions_White_Paper_08Q3.pdf," *White Paper,* 2009, Accessed on September 2010.